\documentclass[12pt]{article}
\usepackage{amssymb}

\usepackage[english]{babel}
\usepackage{graphicx,amssymb}

\begin{document}

\pagestyle{empty}

\begin{flushright}
hep-th/0211022\\
$~$\\
November 2002
\end{flushright}

\vspace{1cm}

\begin{center}
{\large \textbf{Group velocity in noncommutative spacetime}}
\end{center}

\vskip1.5 cm

\begin{center}
{\footnotesize \textbf{Giovanni~AMELINO-CAMELIA}, \textbf{Francesco~D'ANDREA}
and \textbf{Gianluca~MANDANICI}}

\medskip

\textit{Dipart.~Fisica, Univ.~Roma ``La Sapienza'', P.le Moro 2, 00185 Roma,
Italy}
\end{center}

\vspace{1.5cm}

\begin{center}
\textbf{ABSTRACT}
\end{center}

{\leftskip=0.6in \rightskip=0.6in The realization that forthcoming
experimental studies, such as the ones planned for the GLAST space
telescope, will be sensitive to Planck-scale deviations from Lorentz
symmetry has increased interest in noncommutative spacetimes in which this
type of effects is expected. We focus here on $\kappa$-Minkowski spacetime,
a much-studied example of Lie-algebra noncommutative spacetime, but our
analysis appears to be applicable to a more general class of
noncommutative spacetimes. A technical controversy which has significant
implications for experimental testability is the one concerning
the $\kappa$-Minkowski relation between group velocity and momentum.
A large majority of studies adopted
the relation $v = dE(p)/dp$, where $E(p)$ is
the $\kappa$-Minkowski dispersion relation, but recently some authors
advocated alternative formulas.
While in these previous studies the relation between
group velocity and
momentum was introduced through \textit{ad hoc} formulas, we
rely on a direct analysis of wave propagation
in $\kappa$-Minkowski. Our results lead conclusively
to the relation $v = dE(p)/dp$. We also show that the previous
proposals of alternative velocity/momentum relations
implicitly relied on an
inconsistent implementation of functional calculus on $\kappa$-Minkowski
and/or on an inconsistent description of spacetime translations.}

\newpage

%
\baselineskip12pt plus .5pt minus .5pt

\pagenumbering{arabic}

\pagestyle{plain}

\section{Introduction}
Over the last few years there has been a sharp increase in the interest in
experimental investigations of Planck-scale effects (see, \textit{e.g.},
Refs.~\cite{gacEMNS,biller,gacNA1,kifune,gacthresh,qgp,qgreviews}).
In particular,
studies such as the ones planned for the GLAST space telescope~\cite{glast}
would be sensitive to small, Planck-scale suppressed,
deviations from the special-relativistic relation between group velocity and
momentum. Within the framework of Planck-scale spacetime noncommutativity
such modifications of the relation between group velocity and momentum are
often encountered. A noncommutative spacetime which has been considered
extensively in this respect is the $\kappa$-Minkowski
spacetime~\cite{majrue,kpoinap,gacmaj}.
Group velocity in $\kappa$-Minkowski has been
discussed in several studies (see, \textit{e.g.},
Refs.~\cite{kpoinap,gacmaj} and references
therein), under the working assumption that the relation $v = dE(p)/dp$,
which holds in Galilei spacetime and Minkowski spacetime, would also hold in
$\kappa$-Minkowski. This leads to interesting predictions as a result of the
fact that, upon identification of the noncommutativity scale $1/\kappa$
with the Planck length $L_p$, the dispersion relation $E(p)$ that holds
in $\kappa $-Minkowski is characterized by Planck-length-suppressed deviations
from the corresponding relation that holds in Minkowski spacetime. These
deviations are very small, because of the Planck-length suppression, and
they are in agreement with all available data~\cite{gacEMNS,biller,qgp}, but
the mentioned forthcoming experimental studies would be able~\cite{glast} to
test them.

Recently the validity of $v = dE(p)/dp$ in $\kappa$-Minkowski has been
questioned in the studies reported in Refs.~\cite{kowalskiVELOX,tamakiVELOX};
moreover in the study reported in Ref.~\cite{lukieVELOX} the relation
$v =dE(p)/dp$ was considered on the same footing as some alternative relations.
Especially in light of the mentioned plans for experimental studies, this
technical issue appears to be rather significant. We here report progress on
an approach to the study of $\kappa$-Minkowski which was initiated in
Ref.~\cite{gacmaj}. We argue that key ingredients for the correct derivation of
the relation between group velocity and momentum are: (i) a fully
developed $\kappa$-Minkowski differential calculus, and (ii) a proper
description of
energy-momentum in terms of generators of translations. Our analysis
provides support for the adoption of the formula $v = dE(p)/dp$, already
assumed in most of the $\kappa$-Minkowski literature.
We discuss the \textit{ad hoc} assumptions which led to alternatives
to $v = dE(p)/dp$ in
Refs.~\cite{kowalskiVELOX,tamakiVELOX}, and we find that analysis
in Ref.~\cite{tamakiVELOX} was based on erroneous implementation of the
$\kappa$-Minkowski differential calculus, while the analysis in
Ref.~\cite{kowalskiVELOX} interpreted as momenta some quantities
which cannot be
properly described in terms of translation generators.

\section{Preliminaries on $\protect\kappa$-Minkowski
noncommutative spacetime}
In this section we briefly review some results on $\kappa $-Minkowski
spacetime, with emphasis on the aspects that are relevant for the analysis
we present in the following sections. The $\kappa $-Minkowski
spacetime\footnote{We set $\hbar =c=1$.},
\begin{equation}
\lbrack x_{0},x_{j}]
=i {1 \over \kappa} x_{j},\qquad \quad \lbrack x_{j},x_{k}]=0~,
\label{eq:kappaM}
\end{equation}
is an example of Lie-algebra noncommutative spacetime (spacetimes with
commutation relations of the
type $[x_{\mu },x_{\nu }]=iC_{\mu \nu }^{\sigma}x_{\sigma }$).
It came to the attention of the community~\cite{majrue}
primarily because of its role as dual algebra of the momentum
sector of the popular ``bicrossproduct'' $\kappa$-Poincar\'{e} Hopf
algebra\footnote{Other examples of $\kappa $-Poincar\'{e} Hopf algebras
are discussed in Refs.~\cite{kpoinap,kpoinold}.
Besides its relevance for the description of $\kappa$-Minkowski spacetime
(endowed with the differential calculus here described),
the bicrossproduct $\kappa$-Poincar\'{e} Hopf algebra is a particularly
interesting example also because it has been shown~\cite{dsr}
to be suitable for the construction of a symmetry group of deformed finite
Lorentz transformations obtained exponentiating the generators
of its Lorentz sector.}
\begin{eqnarray}
&&[M_{j},M_{k}]=i\epsilon _{jkl}M_{l} ~,~~~
[M_{j},N_{k}]=i\epsilon_{jkl}N_{l}  ~,~~~
[N_{j},N_{k}]=-i\epsilon _{jkl}M_{l}  \nonumber \\
&&[p_{\mu },p_{\nu }]=0~,~~~[M_{j},p_{0}]=0~,~~~[M_{j},p_{k}]=i\epsilon
_{jkl}p_{l}  \nonumber \\
&&[N_{j},p_{0}]=ip_{j}~,~~~[N_{j},p_{k}]=i\delta _{jk}\left\{ \frac{%
1-e^{-2\lambda p_{0}}}{2\lambda }+\frac{\lambda }{2}p^{2}\right\} -i\lambda
p_{j}p_{k}~,  \label{eq:WPcomm}
\end{eqnarray}
where $p_{\mu }=(p_{0},p_{j})$ are the
translation generators, $M_{j}$ are ordinary rotation
generators,
$N_{j}$ are the $\kappa $-deformed boost generators,
and we introduced the convenient notations
$\lambda \equiv \kappa ^{-1}$, $p\equiv |\vec{p}|$.

The algebraic relations (\ref{eq:WPcomm}) are accompanied by coalgebric
structures: the coproducts
\begin{eqnarray}  \label{eq:coprodotti}
& \Delta(p_0)=p_0\otimes 1+1\otimes p_0\qquad\quad \Delta(p_j)=p_j\otimes
1+e^{-\lambda p_0}\otimes p_j &  \nonumber \\
& \Delta(M_j)=M_j\otimes 1+1\otimes M_j\quad\quad \Delta (N_j)=N_j\otimes
1+e^{-\lambda p_0}\otimes N_j+\lambda\epsilon_{jkl}p_k\otimes M_l &
\end{eqnarray}
and the antipodes
\begin{eqnarray}  \label{eq:antipodi}
& S(N_j)=-e^{\lambda p_0}N_j+\lambda e^{\lambda p_0}\epsilon_{jkl}p_k M_l
\quad\quad S(M_j)=-M_j &  \nonumber \\
& S(p_j)=-e^{\lambda p_0}p_j \qquad\qquad S(p_0)=-p_0 ~.
\end{eqnarray}
The ``mass Casimir" (dispersion relation)
of this ``bicrossproduct" $\kappa$-Poincar\'{e} Hopf
algebra is
\begin{equation}  \label{eq:casimir}
\mathcal{C}_{\kappa}(p_0,\vec{p})
=\Big(\frac{2}{\lambda}\sinh\frac{\lambda p_0}{2} \Big)^2
-p^2e^{\lambda p_{0}}
\end{equation}

It is sometimes quickly stated that the
bicrossproduct $\kappa$-Poincar\'{e} Hopf
algebra (\ref{eq:coprodotti}),(\ref{eq:WPcomm}),(\ref{eq:antipodi})
describes the symmetries of the $\kappa$-Minkowski
noncommutative spacetime (\ref{eq:kappaM}).
Actually, the commutation relations (\ref{eq:kappaM}) are
not sufficient to fully characterize a noncommutative geometry:
one must~\cite{majidBOOK} consider the enveloping
algebra\footnote{The
enveloping algebra is the algebra that contains all
(Taylor-expandable) functions
of the noncommutative coordinates.}
generated by (\ref{eq:kappaM})
and introduce a differential calculus on this enveloping algebra.
The natural~\cite{majidBOOK} differential calculus on
the enveloping algebra of $\kappa$-Minkowski is
\begin{equation}
\partial_{j}:f(x):=:\frac{\partial f(x)}{\partial x_{j}}:  \label{derx}
\end{equation}
\begin{equation}
\partial_{0}:f(x):=:\frac{e^{i\lambda {\partial \over \partial t}}
-1}{i\lambda }f(x):=:
\frac{f(\vec{x},t+i\lambda )-f(\vec{x},t)}{i\lambda }:
\label{dert}
\end{equation}
The notation $:f(x):$, conventional in the $\kappa$-Minkowski
literature, is reserved for
time-to-the-right-ordered\footnote{In $\kappa$-Minkowski spacetime
(with its commuting space coordinates and
nontrivial commutation relations only when the time coordinate is involved),
it is easy to see that the natural functional calculus should
be introduced in terms of time-to-the-right-ordered functions or
(the equivalent alternative of)
intuitive rules for time-to-the-left-ordered functions.
In other noncommutative spacetimes the choice of ordering may not be
so obvious.}
functions of the noncommutative coordinates.
The standard symbolism adopted in Eqs.~(\ref{derx})-(\ref{dert})
describes noncommutative differentials in terms of
familiar actions on commutative functions.
The symbols ``$\partial_{j}$" and ``$\partial_{0}$" refer to
elements of the differential calculus on $\kappa$-Minkowski,
while the symbols ``$\partial/\partial  x_{j}$"
and ``$\partial/\partial t$" act as ordinary derivatives on
a time-to-the-right-ordered function of the $\kappa$-Minkowski
coordinates.
For example, Eq.~(\ref{derx})
states that in $\kappa$-Minkowski $\partial_{x}(xt) = t$
and $\partial_{x}[xt^{2}+2i\lambda xt-\lambda ^{2}x+x^{2}t]=
t^{2}+2i\lambda t-\lambda ^{2}+2 x t$,
{\it i.e.} $\partial_{x}$ acts as a familiar $x$-derivative on
time-to-the-right-ordered functions.
Of course, the $\kappa$-Minkowski commutation relations
impose that, if derivatives are standard on
time-to-the-right-ordered functions,
derivatives must be accordingly modified for functions
which are not time-to-the-right ordered.
For example, since $\partial_{x}(xt) = t$ and $\partial_{x}(x) = 1$
(the functions $xt$ and $x$ are time-to-the-right ordered),
also taking into account the  $\kappa$-Minkowski
commutation relation $xt=tx-i\lambda x$, one can obtain
the $x$-derivative
of the function $tx$, which must be
given by $\partial_{x}(tx) = t + i\lambda$.
Similarly, one finds
that $\partial_{x}[t^{2}x+x^{2}t]=t^{2}+2i\lambda t-\lambda ^{2}+2 x t$
(in fact, using the $\kappa$-Minkowski commutation relations
one finds that $t^{2}x+x^{2}t = xt^{2}+2i\lambda xt-\lambda ^{2}x+x^{2}t]$).

The time derivative described by Eq.~(\ref{dert})
has analogous structure, with the only difference that
the special role of the time coordinate in the structure
of $\kappa$-Minkowski spacetime forces~\cite{majidBOOK} one to introduce
an element of discretization in the time direction:
the time derivative of
time-to-the-right-ordered functions
is indeed standard (just like the $x$-derivative of
time-to-the-right-ordered functions is standard),
but it is a standard $\lambda$-discretized derivative
(whereas the $x$-derivative of
time-to-the-right-ordered functions is a standard continuous derivative).

It is only once (the enveloping algebra of) $\kappa$-Minkowski
is equipped with this differential calculus
that one can address physically-meaningful questions, such
as the ones concerning a description of the symmetries
of  $\kappa$-Minkowski. And it is at this level that the duality
between $\kappa$-Minkowski and the
bicrossproduct $\kappa$-Poincar\'{e} Hopf
algebra (\ref{eq:coprodotti}),(\ref{eq:WPcomm}),(\ref{eq:antipodi})
is established.
In fact, from the coalgebric structure
of $\kappa $-Poincar\'{e} one can actually reconstruct
the $\kappa $-Minkowski spacetime through the inner product
\begin{equation}
\Big<\frac{(i\omega )^{n_{0}}}{n_{0}!}\frac{(-ik_{1})^{n_{1}}}{n_{1}!}
\frac{(-i k_{2})^{n_{2}}}{n_{2}!}\frac{(-ik_{3})^{n_{3}}}{n_{3}!}%
,x_{1}^{m_{1}}x_{2}^{m_{2}}x_{3}^{m_{3}}t^{m_{0}}\Big>
=\delta _{n_{0}m_{0}}\delta _{n_{1}m_{1}}\delta _{n_{2}m_{2}}\delta
_{n_{3}m_{3}}
\label{eq:inner}
\end{equation}
which pairs the two algebras.

This duality between $\kappa$-Minkowski
and bicrossproduct $\kappa$-Poincar\'{e} is
also describable as a covariance (in Hopf-algebra sense)
of $\kappa$-Minkowski under bicrossproduct $\kappa$-Poincar\'{e} actions.
Let $w$ be an element of $U(so_{1,3})$
(a function of boosts and rotations) and $\chi$
a function of the momenta,
then the (left) adjoint action is defined
by\footnote{We introduce $w_{(1)}$ and $w_{(2)}$ in the sense
of the ``Sweedler notation" for the coproduct:
$\Delta(w)=\sum_i w_{(1)}^i\otimes w_{(2)}^i\equiv w_{(1)}\otimes w_{(2)}$
(we omit index and summation symbol).}
\[
w\triangleright \chi =w_{(1)}\chi S(w_{(2)})
\]
If $w\in so_{1,3}$ is a linear combination of boost and rotation generators,
from (\ref{eq:coprodotti}) and (\ref{eq:antipodi})
it follows that the action
on a function of the momenta simply reduces to the commutator:
\begin{equation}
M_{i}\triangleright \chi =[M_{i},\chi ]\qquad \quad N_{i}\triangleright \chi
=[N_{i},\chi ]  \label{eq:WsuP}
\end{equation}
and, by construction, satisfies the condition of covariance
(in the Hopf-algebra sense)
\begin{equation}
w\triangleright (\chi \chi ^{\prime })=(w_{(1)}\triangleright \chi
)(w_{(2)}\triangleright \chi ^{\prime }),  \label{eq:cov}
\end{equation}
which, for the action of a generator of $U(so_{1,3})$, is the familiar
Leibniz rule.

The action of an element of the Lorentz sector on the coordinates is
implicitly defined through the relation:
\begin{equation}
\left\langle f(p),w\,\triangleright :\!g(x)\!:\right\rangle =\left\langle
S(w)\triangleright f(p),:\!g(x)\!:\right\rangle   \label{eq:WsuX}
\end{equation}
which amounts to stating that a finite transformation described
by a grouplike element $g$ be self-adjoint with respect to the
inner product ``$\left\langle
,\right\rangle $''\footnote{If $g$ is a \emph{grouplike} element,
{\it i.e.} $\Delta (g)=g\otimes g,$ then $%
S(g)=g^{-1}$ so that $\left\langle g\triangleright {}f(p),g\,\triangleright
:\!g(x)\!:\right\rangle =\left\langle f(p),(g^{-1}g)\,\triangleright
:\!g(x)\!:\right\rangle =\left\langle f(p),:\!g(x)\!:\right\rangle .$ }.
This guarantees the covariance of $\kappa $-Minkowski in the sense of the
Hopf algebras.

By using (\ref{eq:WsuP}) and (\ref{eq:WsuX}) one obtains
\begin{eqnarray}
M_j \, \triangleright :\!f(x)\!: &=&  -i\epsilon_{jkl}x_k \partial_l :\! f(x)\!:
\\ N_j\,\triangleright :\!f(x)\!: &=&  \left[ix_0\partial_j+ix_j\partial_0
- \frac{\lambda}{2}x_j\partial_\mu\partial^\mu\right] :\! f(x)\!:
\end{eqnarray}
which for the generators of $\kappa $-Minkowski means
\begin{equation}
M_{j}\triangleright x_{0}=0\qquad \quad M_{j}\triangleright x_{k}=i\epsilon
_{jkl}x_{l}\qquad \quad N_{j}\triangleright x_{0}=ix_{j}\qquad \quad
N_{j}\triangleright x_{k}=i\delta _{jk}x_{0}  \label{eq:WsuXrl}
\end{equation}

The canonical (left) action of $p_{\mu }$ on the coordinate space is
\begin{equation}
p_{\mu }\triangleright :f(x):=:-i\frac{\partial }{\partial x^{\mu }}f(x):
\end{equation}
so that the finite trasformation is a simple translation:
\begin{equation}
e^{iap}\triangleright :f(x):=:f(x+a):
\end{equation}

A central role in the $\kappa$-Minkowski functional calculus is played by
the ordered exponentials:
\begin{equation}
e^{-i \vec{q} \vec{x}} e^{i q_0 t} ~,  \label{eq:app}
\end{equation}
where $\{q_j,q_0\}$ are four real numbers and $\{x_j,t\}$
are $\kappa$-Minkowski coordinates.
These ordered exponentials enjoy a simple property
with respect to the generators $p_{\mu }$ of translations
of the $\kappa$-Minkowski coordinates:
\begin{equation}
\left\langle p_{\mu },
e^{-i \vec{q} \vec{x}} e^{i q_0 t} \right\rangle=q_{\mu } ~.
\label{nocite}
\end{equation}
We also note that, using the $\kappa$-Minkowski commutation relations,
one finds the relation
\begin{equation}
e^{-i \vec{q} \vec{x}} e^{i q_0 t}=\exp \left( i q_0 t- i \vec{q} \vec{x}
\frac{\lambda q_0 }{1-e^{-\lambda q_0 }}\right)  \label{eq:appb}
\end{equation}
which turns out to be useful in certain applications.

The ordered exponentials $e^{-i\vec{q}\vec{x}}e^{i\omega t}$ also play the
role of plane waves in $\kappa $-Minkowski since on the mass-shell
({\it i.e.} $\mathcal{C}_{\kappa }(q_0,\vec{q})=M^{2}$)
they are solutions~\cite{gacmaj} of the
relevant wave (deformed Klein-Gordon) equation:
\begin{equation}
(\square -M^{2}) \left[ e^{-i \vec{q} \vec{x}} e^{i q_0 t} \right]
=0  \label{waveeq}
\end{equation}
where $\square =\partial _{\mu }\partial ^{\mu }L^{-1}$ is
the $\kappa$-deformed D'Alembert operator,
properly defined~\cite{majidBOOK,gacmaj}
in terms of the so-called ``$\kappa$-Minkowski shift operator" $L$
\[
\;L:\!f(\vec{x},t)\!:
=e^{-\lambda p_{0}}\triangleright :\!f(\vec{x},t)\!:\;
=\;:\!f(\vec{x},t+i\lambda )\!:
\]

The ordered exponentials are also the basic ingredient of the Fourier theory
on $\kappa $-Minkowski. This Fourier theory~\cite{majidBOOK}
is constructed in terms of the canonical
element $\sum_{i}{e_{i}\otimes f^{i}}$, where $\{e_{i}\}$ and $\{f^{j}\}$
are dual bases, which satisfy
the relation $<e_{i},f^{j}>\,\,=\delta _{i}^{j}$.
On the basis of (\ref{eq:inner}) one finds that
the canonical element is
\begin{equation}
\psi _{(q_0 ,\vec{q})}(t,\vec{x})=\sum_{n_{0},n_{1},n_{2},n_{3}}^{0,%
\infty }\frac{(-iq_{1}x_{1})^{n_{1}}}{n_{1}!}
\frac{(-iq_{2}x_{2})^{n_{2}}}{n_{2}!}
\frac{(-iq_{3}x_{3})^{n_{3}}}{n_{3}!}\frac{(i q_0 t)^{n_{0}}}{n_{0}!%
}=e^{-i\vec{k}\vec{x}}e^{i\omega t} \label{eq:scan}
\end{equation}
The canonical element (\ref{eq:scan}) retains the notable feature that, if
we define the transform $\tilde{f}(q)$ of an ordered function $:\!f(x)\!:$
through
\[
:\!f(x)\!:\;=\int \tilde{f}(q)\,e^{-i\vec{q}\vec{x}}e^{iq_0 t}\,
\frac{e^{3\lambda q_{0}}d^{4}q}{(2\pi )^{4}}~,
\]
the choice of the integration measure $e^{3\lambda q_{0}}$ and the
definition (\ref{eq:WsuX}) of the actions of boosts/rotations on the
coordinates guarantee that
\[
w\,\triangleright :\!f(x)\!:\;=\int \left( S(w)
\triangleright \tilde{f}(q)\right) e^{-i\vec{q}\vec{x}}e^{i q_0 t}
\,\frac{e^{3\lambda q_{0}}d^{4}q}{(2\pi )^{4}}
\]
for each $w\in U(so_{1,3})$. This is a relevant property because it implies
that under a finite transformation both $:\!f\!:$ and $\tilde{f}$ change,
but they remain connected by the Fourier-transform relations. The action of
a transformation on the $x$ is equivalent to the inverse transformation on
the $q$. This is exactly what happens in the classical-Minkowski
case ($\lambda =0$), through the simple relation
\[
f(x)\mapsto f_{\Lambda }(x)=\int \!\tilde{f}(\Lambda ^{-1}q)\,e^{iqx}\,
\frac{d^{4}q}{(2\pi )^{4}}=f(\Lambda x) ~.
\]
In $\kappa $-Minkowski the action of boosts does not allow
description in terms of a matrix $\Lambda_{\mu }^{\nu}$,
but it is still true that the action of a transformation
on the $x$ is equivalent to the ``inverse transformation" on the $q$
(where, of course, here the ``inverse transformation" is described
through the antipode).

\section{Group velocity in $\protect\kappa $-Minkowski noncommutative
spacetime}
In this section we first briefly discuss the familiar relation between group
velocity and dispersion relation in classical commutative spacetime, then we
use some of the tools reviewed in the previous section to derive the
corresponding relation that holds in $\kappa$-Minkowski noncommutative
spacetime.

\subsection{Group velocity in commutative spacetimes}
Both in theories in Galilei spacetime and in theories in Minkowski spacetime
the relation between the physical velocity of signals (the group velocity of
a wave packet) and the dispersion relation is governed by the formula
\begin{equation}
v=\frac{dE}{dp}~,  \label{veloxrule}
\end{equation}
in components
\begin{equation}
v_{j}\equiv \frac{dx_{j}}{dt}=\frac{\partial E}{\partial p_{j}}=\frac{p_{j}}{%
p}\frac{\partial E}{\partial p}~.  \label{Vrel}
\end{equation}
This is basically a result of the fact that our theories in
Galilei and Minkowski spacetime
admit Hamiltonian formulation. In classical mechanics this leads directly to
\begin{equation}
\frac{dx_{j}}{dt}=\frac{\partial H(p)}{\partial p_{j}}~.  \label{Vhamil}
\end{equation}
In ordinary quantum mechanics $\vec{x}$ and $\vec{p}$ are described in terms
of operators that satisfy the
commutation relations $[p_{j},x_{k}]=i\delta_{jk}$,
and in the Heisenberg picture the time evolution for the position
operator is given by
\[
\frac{dx_{j}(t)}{dt}=i[x_{j}(t),H]
\]
Since $x_{j}\rightarrow \partial /\partial p_{j}$ and, again, $H\rightarrow
E(p)$, also in ordinary quantum mechanics one finds $v=dE/dp$ (but in
quantum mechanics $v_{j}$ is the operator $dx_{j}/dt$ and the group-velocity
relation strictly holds only for expectation values).

Given a spacetime, the concept of group velocity can be most naturally
investigated in the study of the propagation of waves. It is useful to
review that discussion briefly. For simplicity we consider
a classical 1+1-dimensional Minkowski spacetime.
We denote by $\omega$ the frequency of the wave and
by $k(\omega)$ the wave number of the wave. [Of course, $k(\omega)$
is governed by the dispersion relation, by the mass Casimir of
the classical Poincar\'{e} algebra.]
A plane wave is described by the
exponential $e^{i\omega t-ikx}$. A wave packet is the Fourier
transform of a function $a(\omega)$ which is nonvanishing in a limited
region of the spectrum ($\omega _{0}-\Delta $, $\omega _{0}+\Delta $ ):
\[
\Psi _{(\omega _{0},k_{0})}(t,x)=\int_{\omega _{0}-\Delta }^{\omega
_{0}+\Delta }a(\omega )e^{i\omega t-ikx}d\omega ~.
\]

The information/energy carried by the wave will travel
at a sharply-specified
velocity, the group velocity, only if $\Delta \ll \omega _{0}$.
It is convenient to write the wave-packet
as $\Psi _{(\omega_{0},k_{0})}(t,x)=A(t,x)e^{i\omega_{0}t-ik_{0}x}$,
from which the definition of the wave amplitude $A(t,x)$ follows:
\begin{equation}
A(t,x)=\int_{\omega _{0}-\Delta }^{\omega _{0}+\Delta }a(\omega )e^{i(\omega
-\omega _{0})t-i(k-k_{0})x}d\omega \approx \int_{\omega _{0}
-\Delta}^{\omega _{0}+\Delta }a(\omega )e^{i(\omega -\omega _{0})\left( t
- \left[\frac{dk}{d\omega}\right]_0 x\right) }d\omega  \label{eq:appr}
\end{equation}

The wave packet is therefore the product of the plane-wave factor $%
e^{i\omega _{0}t-ik_{0}x}$ and the wave amplitude $A(t,x)$. One can
introduce a ``phase velocity'', $v_{ph}=\omega _{0}/k_{0}$, associated with
the plane-wave factor $e^{i\omega _{0}t-ik_{0}x}$, but there is no
information/energy that actually travels at this velocity (this ``velocity''
is a characteristic of a pure phase, with modulus $1$ everywhere). It is the
wave amplitude $A(t,x)$ that describes the time evolution of the
energy/information actually carried by the wave packet. From (\ref{eq:appr})
we see that the wave amplitude stiffly translates
at velocity $v_{g}=d\omega_{0}/dk_{0}$,
the group velocity. In terms of the group velocity and the
phase velocity the wave packet can be written as
\[
\Psi _{(\omega _{0},k_{0})}(t,x)=e^{ik_{0}(v_{ph}t-x)}
\int_{\omega_{0}-\Delta }^{\omega _{0}+\Delta }a(\omega )
e^{i(\omega -\omega _{0})(t-x/v_{g})}d\omega ~.
\]

In ordinary Minkowski spacetime the group velocity and the phase velocity
both are 1 (in our units) for photons (light waves) travelling in vacuum.
For massive particles or massless particles travelling
in a medium $v_{ph}\neq v_{g}$.
The causality structure of Minkowski spacetime guarantees
that $v_{g}\leq 1$, whereas, since no information actually travels with the
phase velocity, it provides no obstruction for $v_{ph}>1$.

\subsection{Group velocity in $\protect\kappa$-Minkowski}
The elements of $\kappa$-Minkowski functional analysis we reviewed in
Section~2 allow us to implement a consistent deformation of the
analysis that applies in commutative Minkowski spacetime, here reviewed in
the preceding subsection. In order to present specific formulas we adopt the
$\kappa$-Minkowski functional analysis based on time-to-the-right-ordered
noncommutative functions, but the careful reader can easily verify that the
same result for the group velocity is obtained adopting the time-to-the-left
ordering prescription.

We are little concerned with the concept of phase velocity
(which is not a physical velocity).
In this respect we just observe that the phase velocity should
be a property of the $\kappa$-Minkowski plane wave
\begin{equation}
\psi _{(\omega ,\vec{k})}=e^{-i\vec{k}\vec{x}}e^{i\omega t}~,  \label{pva}
\end{equation}
and, since the $\kappa$-Minkowski calculus is structured in such
a way that the properties of time-to-right-ordered functions
are just the ones of the corresponding commutative function,
this suggests that the relation
\begin{equation}
v_{ph}=\frac{\omega }{k}  \label{pvc}
\end{equation}
should be valid.

But let us focus on the more significant (physically meaningful)
analysis of group velocity. Our starting point is
the wave packet
\[
\Psi _{(\omega _{0},\vec{k}_{0})}=\int e^{-i\vec{k}{\cdot}
\vec{x}}e^{i\omega t}d\mu .\label{kmpack}
\]
In this equation (\ref{kmpack}) for simplicity we denote with $d\mu $ an
integration measure which includes the spectrum of the packet. In fact, the
precise structure of the wave packet is irrelevant for the analysis of the
group velocity: it suffices to adopt a packet which is centered at
some $(\omega _{0},\vec{k}_{0})$ (with $(\omega _{0}$ and $\vec{k}_{0})$
related through Eq.~(\ref{eq:casimir}),
the dispersion relation, the mass Casimir,
of the classical Poincar\'{e} algebra)
and has support only on a relatively small
neighborhood
of $(\omega _{0},\vec{k}_{0})$, \textit{i.e.}
$\omega_{0}-\Delta \omega \leq \omega \leq \omega _{0}+\Delta \omega $
and $\vec{k}_{0}-\Delta \vec{k}\leq \vec{k}\leq \vec{k}_{0}+\Delta \vec{k}$.

Next, in order to procede just following the same steps of the familiar
commutative-spacetime case, we should factor out of the integral a ``pure
phase'' with frequency and wavelength fixed by the wave-packet
center: $(\omega _{0},\vec{k}_{0})$. Consistently with the nature of the
time-to-the-right-ordered functional calculus the phase $e^{i k_{0}x}$ will
be factored out to the left and the phase $e^{-i\omega _{0}t}$ will be
factored out to the right:
\begin{equation}
\Psi_{(\omega _{0},\vec{k}_{0})}=e^{-i\vec{k}_{0}{\cdot}
\vec{x}}\left[ \int e^{-i\Delta \vec{k}{\cdot}\vec{x}}
e^{i\Delta \omega t}d\mu \right] e^{i\omega _{0}t}\!
\label{eq:pacchetto}
\end{equation}
This way to extract the phase factor preserves the time-to-the-right-ordered
structure of the wave $\Psi_{(\omega _{0},\vec{k}_{0})}$, and therefore,
also taking into account the role that time-to-the-right-ordered functions
have in the $\kappa$-Minkowski calculus,
should allow an intuitive analysis of its properties.

From (\ref{eq:pacchetto}) one recognizes the  $\kappa$-Minkowski
group velocity as
\begin{equation}
v_{g}=\lim_{\Delta \omega \rightarrow 0}\frac{\Delta \omega }{\Delta k}
=\frac{d\omega }{dk}~,  \label{gva}
\end{equation}
just as in Galilei and Minkowski spacetime.
Just as one does in commutative Minkowski spacetime, the integral can be
seen as the amplitude of the wave, the group velocity $v_{g}$
is the velocity of translation of this wave amplitude, which
be meaningfully introduced only in the limit of
narrow packet (small $\Delta \omega $ and $\Delta \vec{k}$).

Notice that
\begin{equation}
e^{-i\Delta \vec{k}\vec{x}}e^{i\Delta \omega t}
=\exp \Big(i\Delta \omega t
-i\Delta \vec{k}\vec{x}
\frac{\lambda \Delta \omega }{1-e^{-\lambda \Delta \omega }}\Big)
~,  \label{pvf}
\end{equation}
and
\begin{equation}
\left[\exp \Big(i\Delta \omega t
-i\Delta \vec{k}\vec{x}
\frac{\lambda \Delta \omega }{1-e^{-\lambda \Delta \omega }}\Big)
\right]_{\Delta \omega \rightarrow 0} =
\exp \Big(i\Delta \omega t -i\Delta \vec{k}\vec{x}\Big)
~,  \label{pvfbis}
\end{equation}
and therefore the evaluation of
the velocity of translation of this wave amplitude
turns out to be independent of the way in which the
exponentials are arranged (but this is an accident due to the
fact that for small $\Delta \omega $ and $\Delta \vec{k}$
one finds that $[e^{-i\Delta \vec{k}\vec{x}},e^{i\Delta \omega t}]=0$.


\section{Comparison with previous analyses}
Because of the mentioned interest in the phenomenological
implications~\cite{gacEMNS,biller,qgp,glast},
the introduction of group velocity
in $\kappa$-Minkowski has been discussed
in several studies. In the large majority of
these studies the concept of group velocity was not introduced
constructively (it was not a {\underline{result}} obtained in a full
theoretical scheme: it was just introduced through
an {\underline{{\it ad hoc}}} relation).
This appeared to be harmless since the {\it ad-hoc} assumption
relied on the validity of the relation $v_g = dE/dp$, which holds in Galilei
spacetime and Minkowski spacetime (and for which the
structure of $\kappa$-Minkowski appears to pose no obstacle).

Taking as starting point the approach to $\kappa$-Minkowski
proposed in Ref.~\cite{gacmaj}, we have here shown through
a dedicated analysis that the validity of $v_g = dE/dp$ indeed follows
automatically from the structure of $\kappa$-Minkowski and of the associated
functional calculus.

At this point it is necessary for us to clarify which erroneous assumptions
led to the claims reported in Refs.~\cite{kowalskiVELOX,tamakiVELOX}, which
questioned the validity of $v_g = dE(p)/dp$ in $\kappa$-Minkowski.

\subsection{Tamaki-Harada-Miyamoto-Torii analysis}
It is rather easy to compare our analysis with the study reported
by Tamaki, Harada, Miyamoto and Torii in Ref.~\cite{tamakiVELOX}.
In fact, Ref.~\cite{tamakiVELOX} explicitly adopted the same approach
to $\kappa $-Minkowski calculus that we
adopted here, with Fourier transform and functional calculus that make
direct reference to time-to-the-right-ordered functions. Also the scheme of
analysis is analogous to ours, in that it attempts to derive the group
velocity from the analysis of the time evolution of a superposition
of plane waves. However,
the $\kappa $-Minkowski functional calculus was applied inconsistently in
Ref.~\cite{tamakiVELOX}: at the stage of the analysis were one should
factor out the phases  $e^{-i\vec{k}_{0}{\cdot} \vec{x}}$
and $e^{i\omega_{0}t}$
from the wave amplitude (as we did in Eq.~(\ref{eq:pacchetto}))
Ref.~\cite{tamakiVELOX} does not proceed consistently with the
time-to-the-right-ordered functional calculus. Of course, as done here,
in order to maintain the time-to-the-right-ordered form of the wave packet
it is necessary to factor out
the phases  $e^{-i\vec{k}_{0}{\cdot} \vec{x}}$
and $e^{i\omega_{0}t}$ respectively to the left and to the right,
as we did here. Instead in Ref.~\cite{tamakiVELOX}
both phases are factored out to the left
leading to a form of the wave packet which is not time-to-the-right ordered.
In turn this leads to the erroneous conclusion
that $v_{g}(k)\neq d\omega(k)/dk$, {\it i.e.}  $v_g(p) \neq dE(p)/dp$.

This inconsistency with the ordering conventions is the key factor that
affected Ref.~\cite{tamakiVELOX} failure to
reproduce  $v_g(p) \neq dE(p)/dp$,
but for completeness we note here also that Ref.~\cite{tamakiVELOX}
leads readers to the erroneous impression that in order to introduce the group
velocity in $\kappa$-Minkowski one should adopt the approximation
\begin{equation}
e^{-i\vec{k}\vec{x}} e^{i\omega t} \sim e^{-i\vec{k}\vec{x} + i\omega t} ~,
\label{jap1}
\end{equation}
for generic values of $\omega$ and $\vec{k}$.
Actually, unless $\omega$ and $\vec{k}$
are very small,
this approximation is very poor: it only holds in zeroth order in the
noncommutativity scale $\lambda$ and therefore it does not describe reliably
the structure of $\kappa$-Minkowski (since it fails already in leading order
in $\lambda$, it does not even reliably characterize the main differences
between classical Minkowski and $\kappa$-Minkowski). As we showed
here there is no need for the approximation (\ref{jap1}) in the analysis of
the group velocity of a wave packet in $\kappa$-Minkowski.

\subsection{$\kappa$-Deformed phase space}
As discussed in the preceding Subsection,
it is very easy to compare our study with the study
reported in Ref.~\cite{tamakiVELOX},
since both studies adopted the same approach.
We must now provide some guidance for the comparison
with the study reported by Kowalski-Glikman in Ref.~\cite{kowalskiVELOX}.
Also this comparison is significant for us since Ref.~\cite{kowalskiVELOX},
like Ref.~\cite{tamakiVELOX},
questioned the validity of the relation $v_{g}=dE(p)/dp$,
which instead emerged from our analysis.

Our approach to $\kappa $-Minkowski, which originates from techniques
developed in Refs.~\cite{majrue,gacmaj}, is profoundly different from the
one adopted in Ref.~\cite{kowalskiVELOX}.
In fact, the differences start off
already at the level of the action
of $\kappa $-Poincar\'{e} generators on $\kappa$-Minkowski coordinates.
The actions we adopted are described in
Section~2. They take a simple form on time-to-the-right ordered functions,
but they do not allow description as a ``commutator action'' on generic
ordering of functions in $\kappa $-Minkowski.
Instead in Ref.~\cite{kowalskiVELOX} the action of
the $\kappa $-Poincar\'{e} generators on $\kappa$-Minkowski coordinates
was introduced in fully general terms as a commutator action.
This would allow to introduce a ``phase-space extension''
of $\kappa$-Minkowski~\cite{kowalskiVELOX}
\begin{equation}
[x_{0},x_{j}]=i\lambda x_{j}\qquad [p_{0},x_{0}]=-i \qquad
[p_{k},x_{j}]=i\delta _{jk}e^{-\lambda p_{0}}\qquad
[p_{j},x_{0}]=[p_{0},x_{j}]=0  \label{pskow}
\end{equation}
Taking this phase space (\ref{pskow}) as starting point, Kowalski-Glikman
then found, after a rather lengthy analysis, that ``massless particles move
in spacetime with universal speed of light" $c$~\cite{kowalskiVELOX},
in conflict with the relation $v_{g}=dE(p)/dp$ and the
structure of the mass casimir (\ref{eq:casimir}).
Kowalski-Glikman argued that this puzzling conflict with the
structure of the mass casimir might be due to a missmatch between
the mass-casimir relation, $E(p,m)$,
and the dispersion relation, $\omega(k,m)$: the puzzle could
be explained~\cite{kowalskiVELOX}
if the usual identifications $k \sim p$ and $\omega \sim E$
were to be replaced by $k \sim p e^{\lambda E}$ and
$\omega \sim sinh(\lambda E)/\lambda + \lambda p^2 e^{\lambda E}/2$.

We observe that the correct explanation of the puzzling result
obtained by Kowalski-Glikman is actually much simpler: the commutator
action (\ref{pskow}) adopted in Ref.~\cite{kowalskiVELOX}, in spite of the
choice of symbols $p_j$,$x_k$, {\underline{cannot}} describe the action
of ``momenta" $p_j$ on coordinates $x_k$. Momenta should generate
translations of the coordinates, which requires that they may be represented
as derivatives of functions of the coordinates,
but the commutator action $[p_{k},x_{j}]=i\delta _{jk}e^{-\lambda p_{0}}$
clearly does not allow to represent $p_{k}$ as a derivative with respect
to the $x_{k}$ coordinate, because of the spurious factor $e^{-\lambda p_{0}}$.
Similarly, those commutation relations do not allow to represent
the $x_{k}$ coordinate  as a derivative with respect to $p_{k}$,
and therefore in a Hamiltonian theory, with Hamiltonian $H$, one would find
\begin{equation}
{\dot{x}_{j}} \sim [x_{j},H] \neq {dH \over dp_j}~,  \label{kowHAMILT}
\end{equation}
and this is basically the reason for
the puzzling result $v_{g}(p) \neq dE(p)/dp$
obtained in Ref.~\cite{kowalskiVELOX}.
Kowalski-Glikman finds a function $v_{g}(p)$ but this function cannot
be seen as describing the relation between velocity and momentum,
since the ``$p$" symbol introduced in (\ref{pskow}) does not generate
translations of coordinates, and therefore  ``$p$" is not
a momentum.

\section{Conclusions}
Plans for experimental searches of a possible dependence
of the group velocity on the Planck scale
are already at an advanced stage~\cite{glast}.
The key motivation for these studies
comes from the idea that Planck-scale (quantum)
structure of spacetime might affect the group-velocity/wavelength
relation. This type of Planck-scale effects is plausible (and in
some cases inevitable) in most quantum-gravity approaches,
including phenomenological models of spacetime foam~\cite{gacEMNS},
loop quantum gravity (see, {\it e.g.}, Ref.~\cite{gampul}),
superstring theory (see, {\it e.g.}, Ref.~\cite{sussphotdisp}),
and noncommutative geometry.
However, a detailed careful description of wave propagation is beyond
the reach of our present technical understanding of
most quantum-gravity scenarios. One noticeable exception
is $\kappa$-Minkowski noncommutative spacetime, which is being
considered as a possible flat-space remnant of quantum properties
of spacetime induced by Planck-scale physics.
The structure of $\kappa$-Minkowski spacetime is simple enough
that, as shown here, a rigorous analysis of wave propagation
is possible. In turn this allows us to provide experimentalists
a definite quantum-spacetime scenario to use as reference
for their sensitivity estimates.

We showed here that the formula $v = dE(p)/dp$, where $E(p)$
is fixed by the $\kappa $-Poincar\'{e} dispersion relation,
holds in  $\kappa$-Minkowski spacetime, just like $v = dE(p)/dp$
holds in classical Galileo and Minkowski spacetimes.
The validity of $v = dE(p)/dp$ in $\kappa$-Minkowski had been largely
expected in the literature, even before our direct analysis,
but such a direct analysis had become more urgent after the appearance
of some recent articles~\cite{kowalskiVELOX,tamakiVELOX}
which had argued in favour of alternatives to  $v = dE(p)/dp$
for $\kappa$-Minkowski.
We have shown that these recent claims were incorrect:
the analysis reported
in Ref.~\cite{tamakiVELOX} was based on an erroneous implementation of the
$\kappa$-Minkowski differential calculus, while the analysis in
Ref.~\cite{kowalskiVELOX} interpreted as momenta some quantities
which cannot be properly described in terms of translation generators.

\end{document}